\documentclass[%
 reprint,
superscriptaddress,
 amsmath,amssymb,
aps,
 ]{revtex4-1}

\usepackage{blindtext}
\usepackage[normalem]{ulem}
\usepackage[dvipsnames]{xcolor}
\usepackage{soul}

\usepackage{amsmath,amssymb,amsfonts}
\usepackage{algorithmic}
\usepackage{graphicx}
\usepackage{textcomp}
\usepackage{xcolor}
\usepackage{pifont}
\usepackage[hidelinks,colorlinks=true,linkcolor=blue,citecolor=blue,urlcolor=black]{hyperref}

\usepackage{cleveref}
\usepackage{flushend}
\usepackage{amssymb,amsmath,amsthm,enumitem}
\usepackage{float}

\begin{document}
\title{Experimental Demonstration of Broadband Reconfigurable Mechanical Nonreciprocity}

\author{Amin Mehrvarz}
\thanks{A.M. and M.J.K contributed equally to this work.}
\affiliation{Department of Mechanical and Industrial Engineering, Northeastern University, Boston, MA, 02115, USA.}
\author{Mohammad Javad Khodaei}
\thanks{A.M. and M.J.K contributed equally to this work.}
\affiliation{Department of Mechanical and Industrial Engineering, Northeastern University, Boston, MA, 02115, USA.}
\author{Amir Darabi}
\email{amirdarabi@gatech.edu}
\affiliation{Woodruff School of Mechanical Engineering, Georgia Institute of Technology, Atlanta, GA, 30332, USA.}
\author{Ahmad Zareei}
\email{ahmad@seas.harvard.edu}
\affiliation{Harvard John A. Paulson School of Engineering and Applied Sciences, Harvard University, Cambridge, MA, 02138, USA.}
\author{Nader Jalili}
\email{njalili@ua.edu}
\affiliation{Department of Mechanical Engineering, University of Alabama, Tuscaloosa, AL, 35487, USA.}
\affiliation{Department of Mechanical and Industrial Engineering, Northeastern University, Boston, MA, 02115, USA.}

\begin{abstract}
Breaking reciprocity has recently gained significant attention due to its broad range of applications in engineering systems. Here, we introduce the first experimental demonstration of a broadband mechanical beam waveguide, which can be reconfigured to represent wave nonreciprocity. This is achieved by using spatiotemporal stiffness modulation with piezoelectric patches in a closed-loop controller. Using a combination of analytical methods, numerical simulations, and experimental measurements, we show that contrary to the conventional shunted piezoelectrics or nonlinearity based methods, our setup is stable, less complicated, reconfigurable, and precise over a broad range of frequencies. Our reconfigurable nonreciprocal system has potential applications in phononic logic, wave diodes, energy trapping, and localization.
\end{abstract}

\maketitle

Reciprocity is a fundamental property of various physical systems, where the transmission of a physical quantity, such as waves, between two points in space is symmetrical. Breaking this reciprocity offers an enhanced control over wave signal transmission and has recently become of interest in many branches of physics such as optics \cite{miri2017optical, sounas2017non}, electronics \cite{dobson1995stability}, thermodynamics \cite{torrent2018nonreciprocal, nakai2019nonreciprocal}, electromagnetism \cite{mahmoud2015all, caloz2018electromagnetic}, acoustics \cite{nassar2020nonreciprocity, fleury2014sound}, and classical mechanics \cite{sugino2020nonreciprocal, marconi2020experimental}. In mechanical systems, breaking wave reciprocity has numerous applications in trapping waves for efficient energy harvesting devices and designing transistors in mechanical logic circuits \cite{cummer2016controlling, cullen1958travelling,hadad2016breaking,felsen1970wave,auld1968signal}.

Breaking reciprocity in mechanical systems can be achieved either by (i) passively employing nonlinear elements in an asymmetric structure or (ii) actively varying material properties in space and time periodically \cite{casimir1945onsager}. Passive methods of breaking reciprocity require high wave amplitudes, and as a result are impractical in compact devices \cite{sugino2020nonreciprocal}. Additionally, since passively designed structures cannot be reconfigured or reprogrammed due to their static design, creating robust devices for broad frequency ranges arises additional complexities \cite{darabi2019broadband, fronk2019acoustic, wu2018metastable, boechler2011bifurcation}. Active metamaterials, on the other hand, are reprogrammable and tunable by leveraging the active spatiotemporal modulations. Such active metamaterials offer an effective platform for breaking reciprocity \cite{riva2019generalized, trainiti2016non, nassar2017modulated}, and as a result, have found many applications in (i) increasing the width of bandgaps \cite{airoldi2011design}, (ii) focusing or redirecting wave propagation \cite{celli2017wave, darabi2018broadband, zareei2018continuous}, (iii) changing the amplitude and phase of transmitted and reflected waves \cite{chen2018programmable}, and (iv) one-way wave blocking and cloaking \cite{ning2019active, darabi2018experimental}.

In active metamaterials, active elements such as magnetoelastic \cite{chen2016design, korivand2021band}, photosensitive \cite{yannopoulos2009photoplastic}, or piezoelectric \cite{airoldi2011design, mehrvarz2019vibration}, are used to modulate physical properties. In the photosensitive and magnetoelastic materials, the stiffness can be varied by changing the magnetic field \cite{chen2016design, korivand2021band} and temperature \cite{yannopoulos2009photoplastic}, respectively. In piezoelectric transducers, however, the stiffness is modulated by connecting these patches to shunted circuits \cite{airoldi2011design, marconi2020experimental} {with} negative capacitance \cite{tateo2015experimental}. Although shunted piezoelectric patches are precise \cite{darabi2020reconfigurable, sugino2020nonreciprocal, trainiti2019time, chen2019nonreciprocal}  and able to function in a wide frequency range, a dramatic change in the equivalent Young's modulus only happens when the system operates very close to the unstable zones of the circuit \cite{airoldi2011design}. As such, the shunted piezoelectric patches are prone to an error {where} a small variation of the applied negative capacitance can make the system unstable and result in a large deviation from desired values \cite{tateo2015experimental}. Additionally, implementing a system with shunted piezoelectric patches is extremely challenging since each piezoelectric element is separately connected to its own independent circuit and this requires working with too many elements and connections. 

Here, we present a closed-loop feedback control system connected to two sets of parallel piezoelectric patches \footnote{more specifically, the PZT-5J type, for convenience, PZT refers to this material in the rest of this manuscript} bonded on a host beam to modulate the beam's stiffness. The controller here continually measures the voltage from one set of PZTs and applies the required analog signals to the other set, and as such, changes the beam's effective stiffness \cite{li2017active, sugino2018design}. Contrary to shunted PZTs with negative capacitance circuits, a closed-loop stiffness modulation system is stable and  precise with reduced complexity and has functionality over a broad range of frequencies. As a result, stiffness modulation with a closed-loop control system offers a robust platform for breaking reciprocity. Here, we first analytically/numerically show that spatiotemporal (or spatial) stiffness modulation in beam results in directional (or total) bandgaps. We then experimentally exhibit the bandgaps using a closed-loop controlled system. We show that different behaviors (i.e., wave transmission, blockage, and nonreciprocal transmission) are all attainable on {the same reconfigurable beam setup} by only tuning the controller's parameters.

\begin{figure}[!t]
    \centering
    \includegraphics[width=3.4in,scale=1]{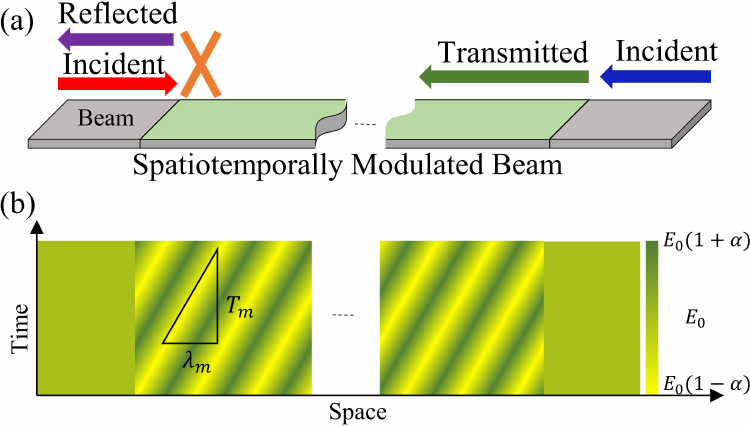}
    \caption{Nonreciprocal wave transmission: (a) Schematic of a beam structure with a spatiotemporally modulated section in the middle. The spatiotemporal modulation results in a nonreciprocal wave propagation allowing wave to propagate from right to left and blocking waves from left to right. (b) The stiffness modulation diagram for the structure is shown in space and time. The spatial/temporal modulation wavelength/period are $\lambda_m$ and $T_m$.}
    \label{fig1}
\end{figure}
\section*{Results}
\textbf{Band-diagrams of a spatio-temporally modulated beam}. We consider elastic wave propagation in a thin beam with linear mass density $\rho_l$, rigidity $E_0$, second moment of inertial $I_0$, and stiffness $D_0 = E_0 I_0$  where the displacement $w$ is governed by the  Euler-Bernoulli equation 
\begin{equation}
    \rho_l\frac{{{\partial }^{2}}w}{\partial {{t}^{2}}}+\frac{{{\partial }^{2}}}{\partial {{x}^{2}}}\left( D(x,t)\frac{{{\partial }^{2}}w(x,t)}{\partial {{x}^{2}}} \right)=0.
    \label{eq1}
\end{equation}

\begin{figure}[!t]
    \centering
    \includegraphics[width=3.4in,scale=1]{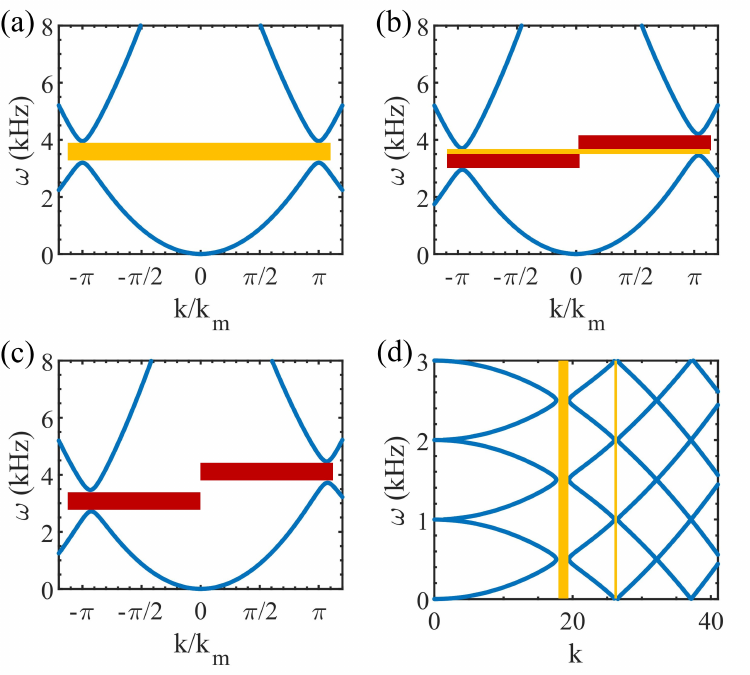}
    \caption{Band diagrams of an aluminium beam ($E=69.9$~GPa and $\rho=2700$~kg.m$^{-3}$) with $1$~mm thickness and $2.5$~cm width. (a) Complete bandgap for only space modulated beam $\alpha_m=0.15$, $k_m=80$~m$^{-1}$, and $\omega_m=0$. (b) Directional and complete bandgaps in the beam with $\alpha_m=0.15$, $k_m=80$~m$^{-1}$, and $\omega_m=400$~Hz. (c) Directional bandgaps in the beam with $\alpha_m=0.15$, $k_m=80$~m$^{-1}$, and $\omega_m=800$~Hz. (d) Wavenumber bandgap with $\alpha_m=0.15$, $k_m=0$, and $\omega_m=800$~Hz (for more information on the effect of $k_m$ and $\omega_m$ on band diagram and bandgaps, readers are referred to supplementary videos A and B).}
    \label{fig2}
\end{figure}

Assuming a spatiotemporal modulation in stiffness with spatial modulation wavelength $\lambda_m$ and temporal modulation period $T_m$, the stiffness is obtained as $ D(x,t)  =D_0 \left( 1+ \alpha_m \cos\left( \omega_m t -k_m x \right)\right)$, where $k_m=2\pi/\lambda_m$ is the spatial modulation wavenumber, $\omega_m=2\pi/T_m$ is the temporal modulation angular frequency,  and $\alpha_m$ is the modulation amplitude (Fig.~\ref{fig1}b). Inserting the modulation stiffness, $D(x,t)$ into Eq.~\eqref{eq1} and taking the Fourier transform, we find the characteristic equation
\begin{align}
    \begin{split}
        &\frac{{{\alpha }_{m}}\gamma }{2}{{\left( k+\left( n-1 \right){{k}_{m}} \right)}^{2}}{{\left( k+n{{k}_{m}} \right)}^{2}}{{{\hat{w}}}_{n-1}} \\ 
        &+\frac{{{\alpha }_{m}}\gamma }{2}{{\left( k+\left( n+1 \right){{k}_{m}} \right)}^{2}}{{\left( k+n{{k}_{m}} \right)}^{2}}{{{\hat{w}}}_{n+1}}\\
        &+\left[ \gamma {{\left( k+n{{k}_{m}} \right)}^{4}}-{{\left( \omega +n{{\omega }_{m}} \right)}^{2}} \right]{{{\hat{w}}}_{n}}=0, 
    \end{split}
    \label{eq2} 
\end{align}
where $\gamma = D_0/\rho_l$, and $\hat{w}_n$ is the $n^\text{th}$ mode of the wave amplitude Fourier transform given by $w(x,t)={{e}^{\mathrm{i} \left( \omega t-kx \right)}}\sum\limits_{p=-\infty }^{+\infty }{{{{\hat{w}}}_{n}}{{e}^{\mathrm{i} p\left( {{\omega }_{m}}t-{{k}_{m}}x \right)}}}$. 
%
%
The band diagram of a spatiotemporally modulated beam is the solution of Eq.~\eqref{eq2} and can be found for different modulation parameters $k_m$, $\omega_m$, and $\alpha_m$. Achieving different behaviors is possible by selecting proper modulation functions and simply changing the controller's parameters accordingly (see Supplementary Information for the effects of density and stiffness modulation with or without a phase difference between them.) To avoid nonlinear effects, we assume a small fixed value of $\alpha_m=0.15$ in our modulation. Initially assuming only a spatial modulation, i.e., $k_m\neq 0, \omega_m=0$, the band diagram exhibits a full bandgap as shown in Fig.~\ref{fig2}a. Next, by introducing and varying temporal modulation, i.e., $\omega_m\neq 0$, we find that the bandgap becomes asymmetrical (Fig.~\ref{fig2}b). Furthermore, increasing the temporal modulation results in a fully asymmetrical bandgap, which has a nonreciprocal wave behavior with unidirectional wave propagation (Fig.~\ref{fig2}c). In the case of only temporal modulation, i.e., $k_m=0, \omega_m\neq 0$, we find that the bandgaps are shifted to the wavenumbers where certain wavelengths are not allowed in the system (see Fig.~\ref{fig2}d) \cite{trainiti2019time}. (A continuous changing of the modulation parameters (i.e., $k_m$ and $\omega_m$) can provide a better insight into spatiotemporal modulation, shown in supplementary movies SV1 and SV2.)

\textbf{Stiffness modulation with PZT patches and closed-loop control}. We use a closed-loop circuit with two parallel PZT patches (one actuator and one sensor) on both sides of the beam to implement spatiotemporal modulation in stiffness. Based on the Euler-Bernoulli theory, the stiffness in a beam is defined $M = -D w_{xx}$, where $M$ is the moment, and $w_{xx}$ is the curvature of the beam. When a PZT actuator and a PZT sensor are attached on each side of a beam extending between $x_l$ and $x_r$, the stiffness relation is modified to \cite{Jalili2010}
\begin{equation}
  D = \frac{ \int_{{{x}_{l}}}^{{{x}_{r}}}{M}dx}{- \int_{x_l}^{x_r} w_{xx} dx}  = D_0 + D_p -  \frac{K_p V_a}{ \int_{x_l}^{x_r} w_{xx} dx},
  \label{eq3}
\end{equation}
where $D_p$ is the stiffness of the PZT patches, $V_a$ is the PZT actuator voltage, and $K_p$ is a factor that depends on PZT parameters (see Supplementary Information for more details). If the actuator's voltage, $V_a$, becomes 
\begin{equation}
    V_a=\frac{\left( D_p -\alpha D_0 \right)}{K_p}  \int_{x_l}^{x_r} w_{xx} dx,
    \label{eq4}
\end{equation}
then the beam's rigidity changes to $D = D_0(1+\alpha)$. Note that $\alpha$ is the modulation amplitude and can depend on the location of PZT, $x$, and also time $t$. To apply the actuator's voltage in Eq.~\eqref{eq4}, the only unknown parameter on the right-hand side is the change in the beam's slope, i.e., $\int_{x_l}^{x_r} w_{xx} dx$, where it can be obtained based on the PZT sensor and its dielectric permittivity \cite{yi2019active}. In a PZT sensor $V_s = C_p \int_{x_l}^{x_r} w_{xx} dx $, where $V_s$ is the sensor's voltage, and $C_p$ is a factor that depends on PZT parameters (see Supplementary Information for the detail). As a result, given sensor voltage $V_s$, if the actuator voltage is applied as
\begin{equation}
    V_a=\frac{\left( D_p -\alpha D_0 \right)}{K_p C_p}V_s, 
    \label{eq5}
\end{equation}
the flexural rigidity of the beam becomes $D= D_0(1+\alpha)$. Modulating the coefficient $\alpha$ using $\alpha = \alpha_m \cos(\omega_m t+\phi)$ for each PZT set (sensor and actuator) and applying the phase $\phi$ based on the location of the PZT set, a spatiotemporal modulation can be achieved. 

\begin{figure}
    \centering
    \includegraphics[width=3.4in,scale=1]{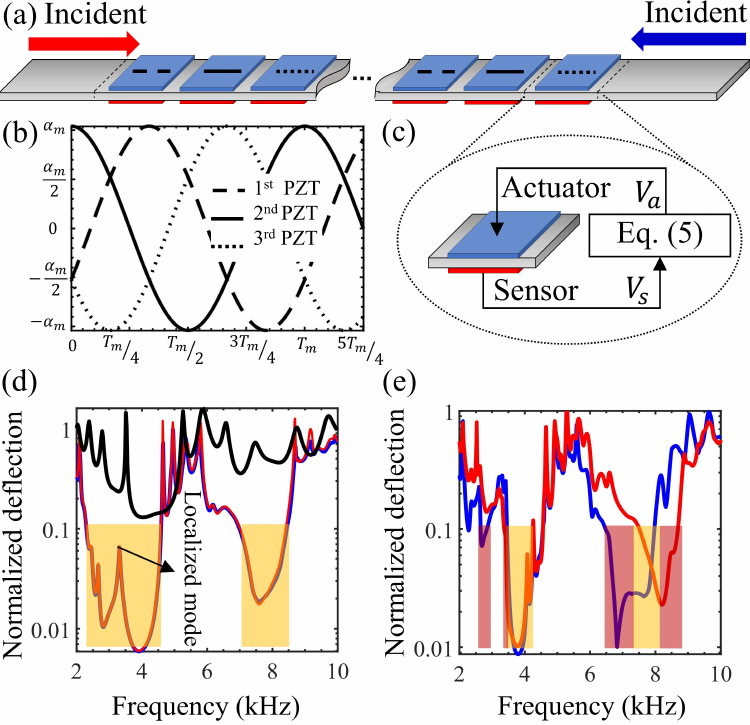}
    \caption{(a) The schematic of the spatiotemporally modulated beam with PZT actuators (blue) and sensors (red) bonded on both sides of the beam. 
    (b) The stiffness modulation coefficient $\alpha_m$ of three consecutive PZT patches over time. Note that the phase difference between consequent patches results in the spatial modulation. (c) The schematic of the closed-loop feedback controller to implement the active modulation. The PZT sensor's voltage is used in Eq.~\eqref{eq5} to obtain actuating voltage, which is applied to the actuator PZT. (d) The numerical transfer function (ratio between transmitted wave to incident wave) without any modulation (black-line) and with spatial modulation with $k_m=80.5$~m$^{-1}$ for waves propagating left-to-right (red line) and waves propagating right-to-left (blue line). Since there is no time modulation, the results are identical for both LR and RL propagations. The yellow boxes indicate the bandgaps. (e) Numerically calculated transmission ratio for spatiotemporally modulated beam with $f_m=400$~Hz and $k_m=80.5$~m$^{-1}$ for waves propagating RL (blue line) and LR (red line). The red boxes indicate the nonreciprocal bandgaps.}
    \label{fig3}
\end{figure}

\textbf{Numerical simulations of spatially and spatio-temporally modulated beam}. We first numerically test the effect of spatiotemporal modulation of the PZTs on an aluminum beam using Finite Element simulations in COMSOL Multiphysics. We consider an aluminum beam with the thickness of $1$~mm, width of $2.54$~mm, and length of $1.5$~m. We further assume twelve pairs of PZT patches perfectly bonded on both sides of the beam. Each PZT covers an area of $21$~mm$\times 21$~mm, thickness $0.55$~mm, and we have a distance of $5$~mm in-between them (see Fig.~\ref{fig3}a). We choose these numbers based on an experimentally feasible setup. A low reflection boundary is used on both sides of the beam to avoid reflecting the wave. The closed-loop parameters are selected so that the stiffness of the piezoelectric actuators vary by $15$ percent, i.e., $\alpha_m=0.15$. The closed-loop feedback controller based on Eq.~\eqref{eq5} is used to modulate the PZT (see Fig.~\ref{fig3}c). The phase difference between subsequent PZTs is set to $\phi=2\pi/3$, which results in three PZT pairs per spatial wavelength $\lambda_m$ (Fig.~\ref{fig3}b). It is to be noted that the equivalent stiffness of each PZT pair, contrary to the shunted circuits, can vary continuously based on the continuous modulation signal (See Fig.~\ref{fig3}b). To test the setup, the wave is initiated on one side of the structure, and measurements are done on both sides of the spatiotemporally modulated section. Finally, a sweep over a range wave frequencies is done to obtain the transfer function (i) in the absence of any modulation, or in the presence of (ii) spatial modulation (i.e., $\omega_m=0, k_m\neq0$), or (iii)  spatiotemporal modulation (i.e., $\omega_m\neq 0, k_m\neq 0$).

In Fig.~\ref{fig3}~d-e, we plot the transfer function, i.e., the ratio of transmitted wave amplitude to incident wave amplitude, in the frequency range $2-10~$kHz for the spatial/spatiotemporal modulation of the beam. Here, the transmission ratio below $r_t\leq 0.1$ is considered blockage because the energy level, which is proportional to the deflection square, drops more than 0.01 \cite{khelif2015phononic}. Note that this transmission level corresponds to $-20dB$ which is commonly considered as bandgap \cite{yi2017frequency,zouari2018flexural, tang2017broadband}. First, in the absence of modulation, no bandgap is observed over the frequency range of interest (black line in Fig.~\ref{fig3}d). In the presence of spatial modulation, however, the transfer function changes and results in identical transfer functions for left-to-right (LR) and right-to-left (RL) wave of propagation (see red and blue lines in Fig.~\ref{fig3}d respectively). Additionally, the results show two identical bandgaps for opposite propagation direction (LR and RL) bounded between $2.5-4.5$~kHz, marked with yellow boxes in Fig.~\ref{fig3}d, and  $7-8.5$~kHz. Interestingly, the first bandgap contains a localized mode at $3.16$~kHz, which dominates the transfer function response, and results in a spike in the transmission ratio in the bandgap. Next, we run the spatiotemporal modulated simulations (i.e., $\omega_m\neq0, k_m\neq0$) and report the results in Fig.~\ref{fig3}e for the wave propagation in LR and RL directions. This figure shows the presence of three directional bandgaps (marked with red rectangles), in which the magnitude of the wave for LR is different from RL. As demonstrated, for two of these bandgaps ($2.65-2.86$~kHz, and $6.46-7.48$~kHz), waves only travel for LR, while for the ($7.97-8.69$~kHz) waves propagate in the opposite direction. Furthermore, bandgaps move to lower frequencies for the RL and higher frequencies for the LR, respectively. Nonetheless, this change is not visible for the first bandgap due to the presence of localized mode in this frequency range.

\begin{figure}[!t]
    \centering
    \includegraphics[width=3.4in,scale=1]{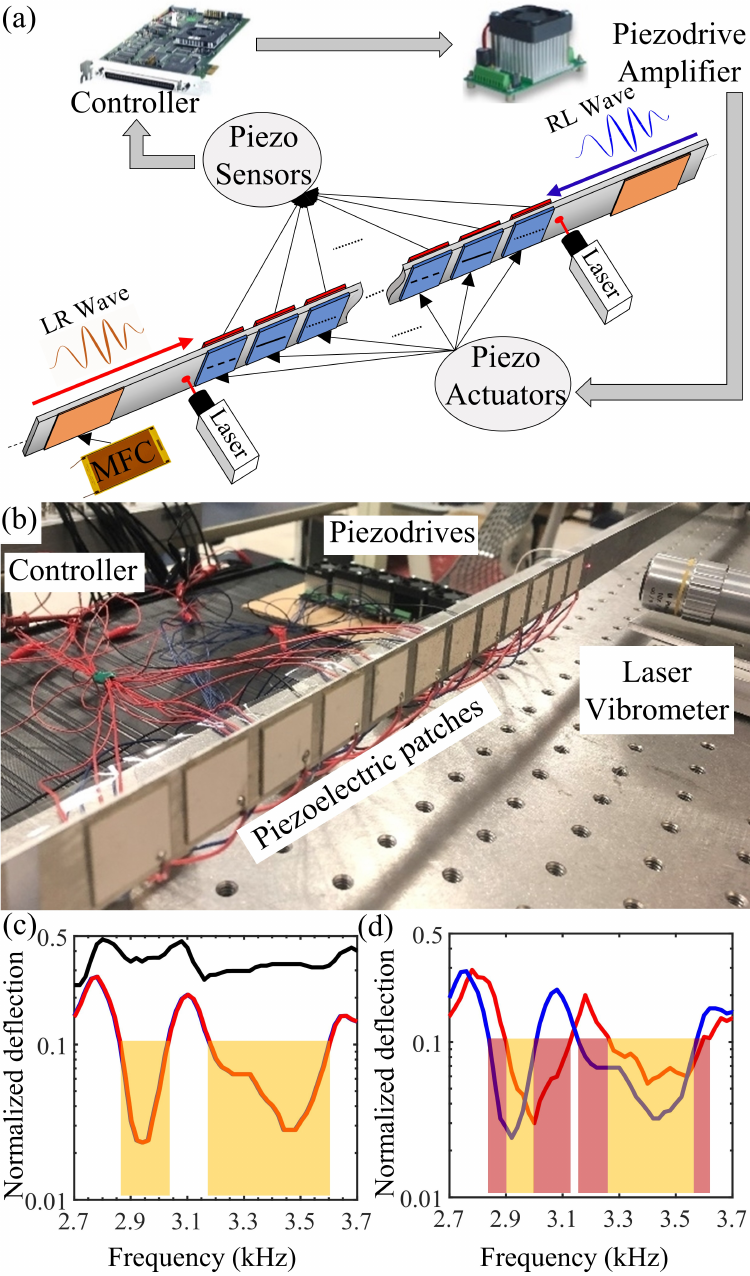}
    \caption{(a) Schematic of the experimental setup showing the beam, absorbing patches,  PZTs sensors and actuators, piezodrives, MFC actuators, controller, and the laser. MFCs on both sides are used to introduce the wave to the system. The wave amplitude is measured before/after the first/last PZTs using a laser vibrometer. (b) Experimental setup. (c) The experimental measurements of transmission ratio without any modulation (solid black line) and with spatial modulation with $k_m=80.5$~m$^{-1}$ for wave propagating LR (red) and RL (blue). Note that the transmission ratio for RL and LR wave are identical. The bandgaps are indicated with the yellow box. (d) Transmission ratio in presence of spatiotemporal modulation with $k_m=80.5$~m$^{-1}$ and $f_m=400$~Hz for LR (red) and RL (blue) waves. The nonreciprocal bandgaps are identified with the red box.}
    \label{fig4}
\end{figure}

\textbf{Experimental frequency response measurement for spatially and spatio-temporally modulated beam}. We implement the time-periodic stiffness modulation of the elastic waveguide beam using an array of PZT actuators and sensors attached on a flexible aluminum beam controlled using closed-loop circuits (Fig.~\ref{fig4}). The aluminum beam has a rectangular cross-section similar to the numerical simulation, and layers of Butyl rubber at its boundaries damp the wave reflection on the structure. The waves are produced using Macro Fiber Composite (MFC) actuators that are bonded on two {ends} of the beam, and a laser vibrometer (Polytec CLV-2544) measures the system's response at a distance of $5$~mm before and after the modulated section (Fig.~\ref{fig4}a). The voltages of PZT sensors are measured using a DS1006 R\&D controller board. The measured voltages are used in the closed-loop circuits in Simulink-MATLAB with reconfigurable parameters of Eq.~\eqref{eq5} (see Supplementary Information for more details). As can be seen in Fig.~\ref{fig4}a, ControlDesk software with DS1006 R\&D controller board applies the control voltages to the piezodrive amplifier, and the outputs are applied to the PZT actuators. See Fig.~\ref{fig4}b for the experimental setup. The experimental results for the transformation ratio without any modulation are shown in Fig.~\ref{fig4}c with a solid black line. 

\section*{Discussion}

Introducing spatial modulation with $k_m= 80.5$~m$^{-1}$, the transmission ratio changes and results in two identical bandgaps for RL and LR wave propagation direction at frequencies $2.86-3.33$~kHz and $3.18-3.6$~kHz (shown with yellow boxes in Fig.~\ref{fig4}c). Note that in the analytical results, this bandgap is continuous from 2.86-3.6~kHz without any break in the middle. However, in the experiments similar to the numerical simulations, this bandgap breaks into two separate bandgaps due to an internal localized mode of the system. Introducing a time modulation in addition to the spatial modulation with $f_m =400$~Hz results in asymmetrical bandgaps that depend on the direction of wave propagation. Figure~\ref{fig4}d shows the presence of four directional bandgaps (red boxes), in which, same as the numerical simulations, the magnitude of the wave for LR is different from RL. Interestingly, two of the bandgaps (specifically $2.84-2.9$~kHz, and $3.16-3.27$~kHz) allow only LR wave propagation (RL nonreciprocal bandgaps), while on the contrary, the other two band gaps (i.e., $3.01-3.13$~kHz and $3.57-3.6$~kHz) allow only RL waves propagation (LR nonreciprocal bandgaps). Moreover, similar to the numerical results, bandgaps move to lower/higher frequencies for the RL/LR bandgaps. It should be noted that two of the directional bandgaps are observed only because the bandgaps are in the vicinity of the interior localized mode. Additionally, we expect the experimental result to defer from the numerics due to the following sources of error, (i) In the numerical simulations, the contacts between the PZTs and beam are assumed to be perfect contacts; however, there is an adhesive layer between the PZTs and aluminum beam that changes the effective thickness of the system but also changes the corresponding density of the system \cite{rabinovitch2002adhesive}. Additionally, (ii) our experiments include structural damping and nonlinear effects, which are ignored in the simulations and theory. Lastly, (iii) including the equipment's error (e.g., signal generator, piezodrive amplifier, and controller) would further defer our experimental result from the numerical simulation.
%


In summary, we experimentally implemented a beam structure to achieve total/nonreciprocal bandgaps. For the first time, an active closed-loop feedback control system was employed using an array of PZTs bonded on both sides of an aluminum host layer to implement the stiffness modulation. At first, we investigated the changing effects of spatiotemporal parameters on the bandgaps. Then, we numerically simulated spatial/spatiotemporal stiffness modulation using the closed-loop system and validated our analytical results. Finally, we experimentally showed that actively controlled stiffness modulation using closed-loop circuits is reconfigurable and can be used for both spatial and spatiotemporal modulation, and has potential applications in waveguides, diodes, phononic logic circuits, or energy localization. 
\vspace{5mm}

\section*{Data availability}
The data that support the plots within this paper and other findings of this study are available from the corresponding author upon request.

\vspace{4mm}
\section*{Acknowledgments}
The authors declare that they have no known competing for financial interests or personal relationships that could have appeared to inﬂuence the work reported in this paper.

\end{document}